# Structural and magnetic characterization of the one-dimensional $S = 5/2$ antiferromagnetic chain system SrMn(VO$_4$)(OH)

Liurukara D. Sanjeewa,[1] V. Ovidiu Garlea,[2] Michael A. McGuire,[3] Colin D. McMillen,[1] Huibo Cao,[2] and Joseph W. Kolis [1*]

[1]*Department of Chemistry and Center for Optical Materials Science and Engineering Technologies (COMSET), Clemson University, Clemson, South Carolina 29634-0973, USA*
[2]*Quantum Condensed Matter Division, Oak Ridge National Laboratory, Oak Ridge, TN 37831, USA*
[3]*Materials Science and Technology Division, Oak Ridge National Laboratory, Oak Ridge, Tennessee, 37831, USA*

**Abstract**

The descloizite-type compound, SrMn(VO$_4$)(OH), was synthesized as large single crystals (1-2mm) using a high-temperature high-pressure hydrothermal technique. X-ray single crystal structure analysis reveals that the material crystallizes in the acentric orthorhombic space group of $P2_12_12_1$ (no. 19), $Z = 4$. The structure exhibits a one-dimensional feature, with [MnO$_4$]$_\infty$ chains propagating along the *a*-axis which are interconnected by VO$_4$ tetrahedra. Raman and infrared spectra were obtained to identify the fundamental vanadate and hydroxide vibrational modes. Magnetization data reveal a broad maximum at approximately 80 K, arising from one-dimensional magnetic correlations with intrachain exchange constant of J/k$_B$ = 9.97(3) K between nearest Mn neighbors and a canted antiferromagnetic behavior below $T_N$ = 30 K. Single crystal neutron diffraction at 4 K yielded a magnetic structure solution in the lower symmetry of the magnetic space group $P2_1$ with two unique chains displaying antiferromagnetically ordered Mn moments oriented nearly perpendicular to the chain axis. The presence of the Dzyaloshinskii–Moriya antisymmetric exchange interaction leads to a slight canting of the spins and gives rise to a weak ferromagnetic component along the chain direction.

## I. INTRODUCTION

Significant interest in one-dimensional (1D) systems has stemmed from their exotic magnetic properties and spin dynamics that are mainly driven by the interplay of quantum fluctuations and thermal fluctuation. These magnetic properties are strongly dependent on the spin values, as well as on existence of single-ion anisotropy [1]. A distinct attention has been initially devoted to 1D systems with small spins (S = 1/2 or 1), but the focus has progressively expanded to systems with larger spins for which even a small inter-chain coupling gives rise to three-dimensional (3D) magnetic order [2-10]. The strong spin correlations within the spin chains of these systems were found to manifest themselves above the long-range order by a broad maximum in the susceptibility versus temperature data [2,3]. Furthermore, neutron scattering experiments revealed that short-wavelength excitations along the chain direction persist into the paramagnetic regime, even at temperatures of about twice the Néel temperature ($T_N$) [7-9]. Systematic studies on chain-based classical-spin systems were carried out more than four decades ago on $[(CH_3)_4N]$ $MnCl_3$ (also known as TMMC) [4-7], $CsMnCl_3 \cdot 2H_2O$ [8,9], and $CsMnBr_3$ [10]. Each of these compounds consists of weakly interacting chains of $Mn^{2+}$ with S=5/2, and magnetic studies revealed striking 1D magnetic behavior.

Over the years, several new classes of 1D dimensional systems have emerged and notable among them are mixed alkaline earth and transition metal silicates, $BaM_2Si_2O_7$ with M = Cu, Co, Mn [11,12] and the pyroxene materials $AMX_2O_6$, where A is an alkali ion, M is a transition metal ion while $X$=Si, Ge [13-15 and references therein]. The interplay of low dimensionality and magnetic frustration in the pyroxenes offered new opportunities to tune the richness of the physic in this family and gave rise to multiferroic properties [14]. Another rich series of new low dimensional materials are those containing tetrahedral oxyanions (e.g., $VO_4^{3-}$, $PO_4^{3-}$, $SO_4^{2-}$) as bridging units. These exhibit an extensive structural versatility and can link magnetic ions in a variety of environments. Thus a wide variety of new structural and magnetic behaviors can be examined. One tetrahedral building block with a wide variety of bridging modes is the vanadate group $(VO_4)^{3-}$. It not only displays a wide range of bridging environments but can also accommodate a broad assortment of metal ions and spin values [16-22]. In addition to its structural versatility the presence of $d$ orbitals on the bridging group introduces the possibility of the modification of the Goodenough-Kanamori-Anderson superexchange rules [23]. In this context, the compounds belonging to the class $AM_2V_2O_8$ (where $A$ = Sr, Ba, and $M$ = Cu, Ni, Co, Mn) have recently attracted attention to study the role of anisotropy, spin-value, and interchain interactions on the 1D magnetic properties [20-22].

Recently we demonstrated that another series of metal vanadates, $Ba_2M(VO_4)_2(OH)$ ($M$ = $V^{3+}$, $Mn^{3+}$, $Fe^{3+}$), with unusual low dimensional magnetic coupling over a range of spin values could be prepared and investigated [24]. This paper describes an extension of these investigations to a new low dimensional vanadate chain containing $Mn^{2+}$, S = 5/2 ions. This system displays pronounced quasi-one-dimensional magnetic characteristics and a distinct phase transition to a long–range canted antiferromagnetic order that involves a symmetry reduction from orthorhombic to monoclinic. In addition, we demonstrate the presence of a Dzyaloshinskii–Moriya antisymmetric exchange interaction that leads to a weak ferromagnetic component along the chain direction.

### II. EXPERIMENTAL DETAILS

Single crystals of $SrMn(VO_4)(OH)$ were obtained by a high temperature hydrothermal method. Reactions were performed in 2.5 in long silver ampoule with outer diameter of 0.25 in. A total of 0.2 g of reactants consisting of SrO (0.0569 g), $Mn_2O_3$ (0.0433 g) and $V_2O_5$ (0.0998 g) were mixed in a molar ratio of 2:1:2 with 0.4 mL of 5 M $K_2CO_3$ water. The loaded silver ampoule was welded and placed in a Tuttle-seal autoclave filled with deionized water to achieve a desirable counter pressure. The autoclave was heated at 580 °C for 6–7 days, and then cooled to room temperature. Brown polyhedral crystals of 1–2 mm were retrieved from silver ampoules by washing with deionized water. The chemical composition of the crystals was verified by energy dispersive X-ray (EDX) methods.

Selected crystals of $SrMn(VO_4)(OH)$ were characterized by X-ray diffraction using a Rigaku AFC-8S diffractometer equipped with graphite monochromated Mo Kα radiation (λ = 0.71073 Å). Diffraction data was collected at room temperature. The structure was solved by direct methods using the SHELX software suite and refined on $F^2$ by full-matrix least squares techniques [25]. All non-hydrogen atoms were refined anisotropically. The hydrogen atom of the asymmetric unit was identified from the difference electron density map, and its position was constrained to prevent unreasonable variations in the O−H bond length.

Vibrational spectroscopic characterization was performed using single crystals of $SrMn(VO_4)(OH)$ mixed with KBr and ground thoroughly before pressing into a translucent pellet. The infrared spectrum was collected on a Nicolet Magna IR Spectrometer 550 in the frequency range from 400 $cm^{-1}$ to 4000 $cm^{-1}$ with a 4 $cm^{-1}$ resolution. Raman scattering on ground single crystals was performed using a 514.5 nm wavelength Ar laser with 90 mW (Innova 200, Coherent) with a camera lens of f/1.2 in a backscattering geometry. Raman scattering was detected by a triple spectrometer (Triplemate 1877, Spex) equipped with a charge-coupled device (CCD) detector (iDUS 420 series, Andor) cooled to 60 °C. For calibration purposes, the Raman spectrum of Indine and the Raman spectrum of a 5:2 mixture of chloroform/bromoform was used. The typical spectrum acquisition time was 10 s.

Temperature and field-dependent magnetic measurements were carried out using a Quantum Design Magnetic Property Measurement System (MPMS). The measurements were carried out on a single crystal specimen of approximately 5.94 mg oriented such that the [MnO$_4$]$_\infty$ chains (i.e. *a*-axis) were aligned either parallel or perpendicular to the applied magnetic field. The temperature dependence of static susceptibility [*M/H*(T)] was measured over a temperature range of 2–300 K for in the applied fields of 1 kOe and 10 kOe. The isothermal magnetization measurements were performed at 2, 5, and 300 K for fields up to 50 kOe.

Single crystal neutron diffraction data were collected using the HB3A four-circle diffractometer at the High Flux Isotope Reactor at Oak Ridge National Laboratory [26]. The measurements were conducted on the same single crystal sample as that subjected to magnetization measurements. The crystal specimen glued on the top of an aluminium rod was loaded in a closed-cycle-refrigerator whose temperature was controlled in the range 4–300 K. For the measurements we used a monochromatic beam with the wavelength 1.542 Å selected by a multilayer-[110]-wafer silicon monochromator, and the scattered intensity was measured using an Anger-camera type detector. The neutron diffraction data were analyzed by the using the FullProf Suite package [27].

**III RESULTS AND DISCUSSION**

*A. Crystal Structure of SrMn(VO$_4$)(OH)*

The detailed structure of SrMn(VO$_4$)(OH) is critical to the understanding of the magnetic structure, so it is dealt with carefully here. The overall structure is similar to the adelite-descloizite mineral type [28]. This group includes a number of member compounds that can be classified according to space groups *P*2$_1$2$_1$2$_1$ or *Pnma*. Both space groups are well-established depending on sample-specific systematic absences. Among many naturally-occurring examples, adelite (CaMg(AsO$_4$)(OH)) crystallize in *P2$_1$2$_1$2$_1$* [29], while descloizite (PbZn(VO$_4$)(OH)) crystallizes in *Pnma* [30]. The systematic absences exhibited by the title compound is consistent with space group *P2$_1$2$_1$2$_1$*, having a Flack parameter of -0.010(6), and the use of software to search for higher symmetry served as confirmation of this selection since no higher symmetry arrangement was appropriate. Though many vanadates of this family crystallize in space group *Pnma*, compounds with slightly smaller divalent cations (Ca$^{2+}$ or Sr$^{2+}$ compared to Pb$^{2+}$, for example) crystallize in space group *P2$_1$2$_1$2$_1$*, and this seems reasonable in the present case. The difference between these arrangements occurs due to displacement of certain oxygen atoms in the structure, though the overall structural motifs are consistent in both space groups.

The structure of SrMn(VO$_4$)(OH) is comprised of one crystallographically distinct [Mn$^{2+}$O$_6$] unit (*oct*-MnO$_6$) and one [V$^{5+}$O$_4$] unit (*tet*-VO$_4$). The *oct*-MnO$_6$ forms [MnO$_4$]$_\infty$ chains along *a*-axis through shared edges. Figure 1(a) shows the partial polyhedral view of the structure along *a*-axis. The *tet*-VO$_4$ unit interconnects Mn−O−Mn chains along the *bc* plane. As shown in Figure 1(b) *oct*-MnO$_6$ share edges through O(1) and O(2), while the *tet*-VO$_4$ unit connects to two adjacent *oct*-MnO$_6$ through axial oxygen atoms. The *tet*-VO$_4$ unit also connects to one of the edge-shared oxygen, O(2) in [MnO$_4$]$_\infty$ chains while another edge-sharing oxygen, O(1), comprises the OH$^-$ group. The detailed crystallographic data of SrMn(VO$_4$)(OH) is given in Table 1. For evaluating the magnetic interactions it is necessary to pay special attention to Mn−O interatomic distances and Mn−O−Mn bond angles. In *oct*-MnO$_6$, Mn−O bond lengths range from 2.046(3) to 2.280(3) Å, comparable to the sum of the Shannon crystal radii (2.230 Å) for a six-coordinate Mn$^{2+}$ and O$^{2-}$ [31]. Further, *oct*-MnO$_6$ is highly distorted by having two short Mn−O bonds and four long bonds. The two shortest Mn−O bonds are equatorial bonds to O(1) which is the OH$^-$ group in the structure and participates in edge-sharing within the chains along with O(2). The chains exhibit Mn−Mn separations of 3.071(6) Å, with Mn−O(1)−Mn angles of 96.69(9)° and Mn−O(2)−Mn angles of 86.68(8)°. The relevant Mn−O interatomic distances and angles are summarized in Table 2. The $V^{5+}$−O bond lengths in the VO$_4$ tetrahedra range from 1.686(3) Å to 1.742(2) Å which is also comparable to sum of the Shannon crystal radii for $V^{5+}$−$O^{2-}$ (1.685 Å). The shortest of these bonds is to O(5), a terminal oxygen atom that is not a part of the edge-sharing transition metal chains. This oxygen atom acts as a hydrogen bond acceptor in the O(1)−H(1)⋯O(5) interaction which connects neighboring chains along the $b$-axis through their hydroxide and vanadate groups (O−H⋯O-V = 2.770 Å, 165.23° in SrMn(VO$_4$)(OH)).

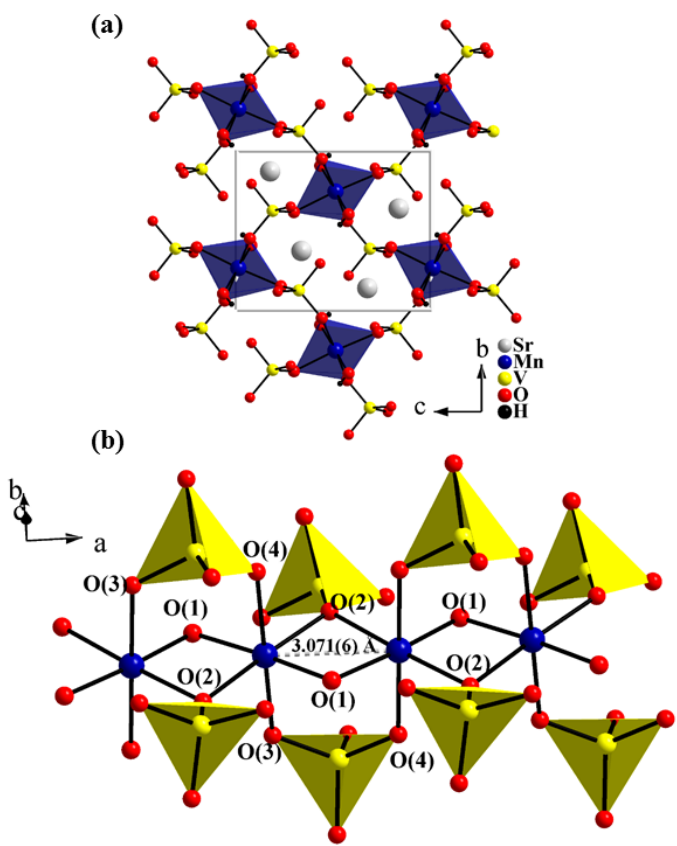

Figure 1. (a) Crystal structure of SrMn(VO$_4$)(OH) viewed along the [100] direction. (b) [MnO$_4$]$_\infty$ chains in SrMn(VO$_4$)(OH) made from edged-shared *oct*-MnO$_6$ via edge-sharing equatorial O(1) and O(2) atoms. The [MnO$_4$]$_\infty$ chains are interconnected by *tet*-VO$_4$ units by chelating axial O(2), O(3) and O(4) atoms in the chain.

**Table 1**
Structural parameters of SrMn(VO$_4$)(OH) at 300 K (s.g. $P2_12_12_1$ (no. 19), Z= 4, a = 6.1322(13) Å, b = 7.7201(19) Å, c = 9.414(2) Å; R1 = 0.0175, wR2 = 0.0405; Flack x = -0.010(6))

| Atom | Wyckoff | x/a | y/b | z/c | Ueq |
|------|---------|-----|-----|-----|-----|
| Sr   | 4a | 0.02489(8)  | 0.86680(5) | 0.83053(5) | 0.00998(18) |
| Mn   | 4a | 0.25420(11) | 0.25901(10)| 0.99443(10)| 0.0089(2)   |
| V    | 4a | 0.98836(13) | 0.88094(9) | 0.18244(8) | 0.0062(2)   |
| O(1) | 4a | 0.0069(6)   | 0.9031(4)  | 0.5724(3)  | 0.0080(7)   |
| H(1) | 4a | 0.036(11)   | 0.989(6)   | 0.516(5)   | 0.025(17)   |
| O(2) | 4a | 0.0084(6)   | 0.0549(4)  | 0.0653(3)  | 0.0096(7)   |
| O(3) | 4a | 0.2297(6)   | 0.8549(5)  | 0.2754(4)  | 0.0120(7)   |
| O(4) | 4a | 0.7622(6)   | 0.8892(5)  | 0.2916(4)  | 0.0110(7)   |
| O(5) | 4a | 0.9406(6)   | 0.7119(4)  | 0.0735(4)  | 0.0153(8)   |

**Table 2**
Selected interatomic distances (Å) and angles (°) for Sr$_2$Mn(VO$_4$)(OH)

| Atoms | Bond lengths & angles |
|-------|----------------------|
| Mn−O(1) | 2.046(3) |
| Mn−O(1) | 2.061(3) |
| Mn−O(2) | 2.193(3) |
| Mn−O(2) | 2.280(3) |
| Mn−O(3) | 2.244(3) |
| Mn−O(4) | 2.254(3) |
| Mn−O(1)−Mn | 96.8(2) |
| Mn−O(2)−Mn | 86.7(1) |

### B. Raman and Infrared Spectra SrMn(VO$_4$)(OH)

Figure 2 shows the infrared and Raman spectra of SrMn(VO$_4$)(OH). As shown in Figure 2a, the sharp band at 3733 cm$^{-1}$ in the IR can be assigned to O-H stretching. This is somewhat high relative to typical OH stretches in these systems [32], and suggests that any hydrogen bonding is somewhat weak. Bands in the range of 1100-600 cm$^{-1}$ are due to the stretching vibrations of the [VO$_4$] group in the structure. Figure 2b displays the Raman spectrum of SrMn(VO$_4$)(OH) in the region of 100-1000 cm$^{-1}$. The spectrum is in good agreement with the Raman spectra of various vanadates with similar structures [33]. The vanadate ions are only approximately tetrahedral and the lower site symmetry of the coordinated vanadates, as well as the acentric crystal structure, causes somewhat more complex vibrational behavior. Previous work on natural descloizite minerals shows peaks at 858 and 849 cm-1 were assigned as $\nu_1$ symmetric stretches. [34] A somewhat less intense peak at 779 was assigned as a $\nu_3$ antisymmetric stretch. Our Raman spectra display similar peaks at 831 783 and cm$^{-1}$. The shape and intensity of the peaks is very similar to the peaks observed for BaAg$_2$Mn(VO$_4$)$_2$, which were assigned as Ag, Ag and Eg stretches respectively. [35] These workers also observed a signficant sensitivity of the frequencies to the nature of the other metal ions in the structure. Peaks at 446 and 338 cm$^{-1}$ can be assigned to the $\nu_1$ and $\nu_3$ bending modes of the [VO$_4$] respectively, as in natural descloizite (Figure 2b inset). [34] Weaker broad peaks below 300 cm$^{-1}$ are probably due to the Sr−O and/or lattice vibrations. A similar, but somewhat simpler Raman spectrum is observed in the present study compared to that of Ba$_2$V(VO$_4$)$_2$(OH) reported previously [24], perhaps reflecting the presence of only one unique vanadate group in SrMn(VO$_4$)(OH), while Ba$_2$V(VO$_4$)$_2$(OH) contains two structurally distinct vanadate groups.

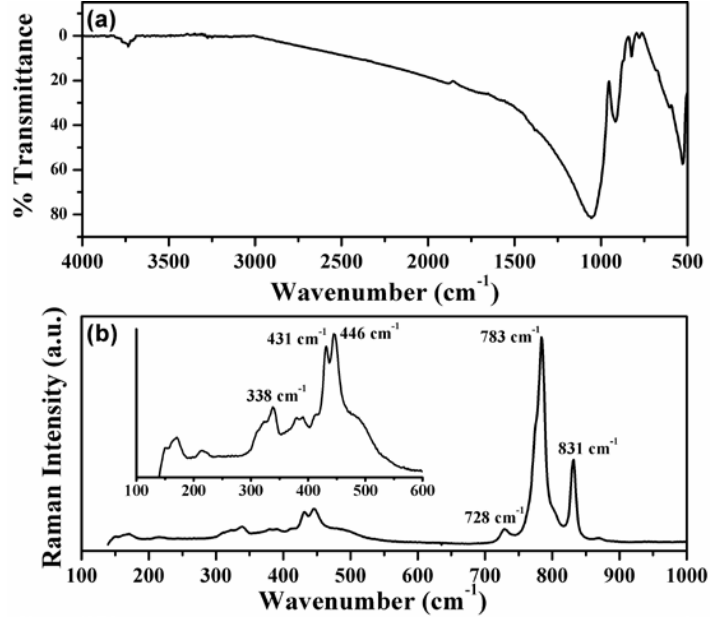

Figure 2. Infrared (a) and Raman (b) spectra of SrMn(VO$_4$)(OH); Figure 2b inset shows an enlarged view of the Raman spectrum of SrMn(VO$_4$)(OH) in the region of 100 to 600 cm$^{-1}$.

## C. Magnetization and heat capacity

The temperature dependence of the static susceptibility, $\chi$ = M/H, measured both parallel and perpendicular to the [MnO$_4$]$_\infty$ chain axis that propagates along a ($H \parallel a$ and $H \perp a$) is displayed in Figure 3a. The susceptibility curves show a broad maximum at approximately 80 K for both parallel ($H \parallel$ a) and perpendicular ($H \perp$ a) directions, suggesting the development of 1D short-range spin correlations. Upon further cooling the susceptibility exhibits an abrupt upturn at about 30 K that indicates the onset of a long-range magnetic ordering. This phase transition is more clearly observed in measurements in a lower magnetic field of 1kOe, shown in the inset of Figure. 3a. A small upturn upon cooling is seen at the lowest temperatures in the data collected at 10 kOe, attributable to a low concentration of spins that remain paramagnetic down to 2 K. Assuming these are tetravalent vanadium (S = ½) gives a concentration of 0.04 per formula unit determined from the Curie law fit to the data below 6 K.

The increase in susceptibility that occurs upon cooling near 30 K suggests a ferromagnetic component to long-range the magnetic order; however, isothermal magnetization curves (Figure 3b) measured below the transition do not show saturating behavior. At 2 K the magnetization reaches a value of only 0.15 $\mu_B$/Mn$^{2+}$ at 50 kOe, much smaller than the full moment of 5 $\mu_B$/Mn$^{2+}$. This indicates that antiferromagnetic order with canting of the moments producing a small net magnetization occurs below $T_N$ = 30 K, which is confirmed by the neutron diffraction results presented below. A canted antiferromagnetic structure is consistent with the divergence of the field cooled (FC) and zero field cooled (ZFC) magnetization data (Figure 3a, inset), the small remanent magnetization (Figure 3b) and hysteretic behavior in the magnetization loop (Figure.3b, inset) measured below the ordering temperature. All of these behaviors are indicative of formation of magnetic domains with uncompensated moments. It is worth noting that the presence of a weak ferromagnetic component has also been observed in the quasi-1D system BaMn$_2$(VO$_4$)$_2$ [[20]. A canted structure may be expected due to antisymmetric Dzyaloshinskii-Moriya (DM) interactions in SrMn(VO$_4$)(OH), since it possess a non-centrosymmetric crystal structure with the space group of $P2_12_12_1$. The change in slope of the 2 K magnetization curve (Figure 3b) near 15 kOe suggests that a spin reorientation is induced at higher fields.

Under the assumption of weak interchain interactions the high-temperature susceptibility above the long-range ordering transition can be well described by a modified Bonner-Fisher model for linear magnetic chains with spin $S = 5/2$ [5,6]:

$$\chi = \chi_0 + \frac{N_A S(S+1)}{3k_B T} g^2 \mu_B^2 \frac{1 + u(K)}{1 - u(K)} \quad (1)$$

$$u(K) = \coth K - \frac{1}{K}; \quad K = \frac{2JS(S+1)}{kT} \quad (2)$$

where, $\chi_0$ is a temperature-independent susceptibility due to Van-Vleck paramagnetism or to diamagnetic contributions, $k_B$ is the Boltzmann constant, $g$ is Landé factor, and $J$ is the antiferromagnetic exchange constant coupling nearest neighbors along the chain and $S = 5/2$ of $Mn^{2+}$ ions.

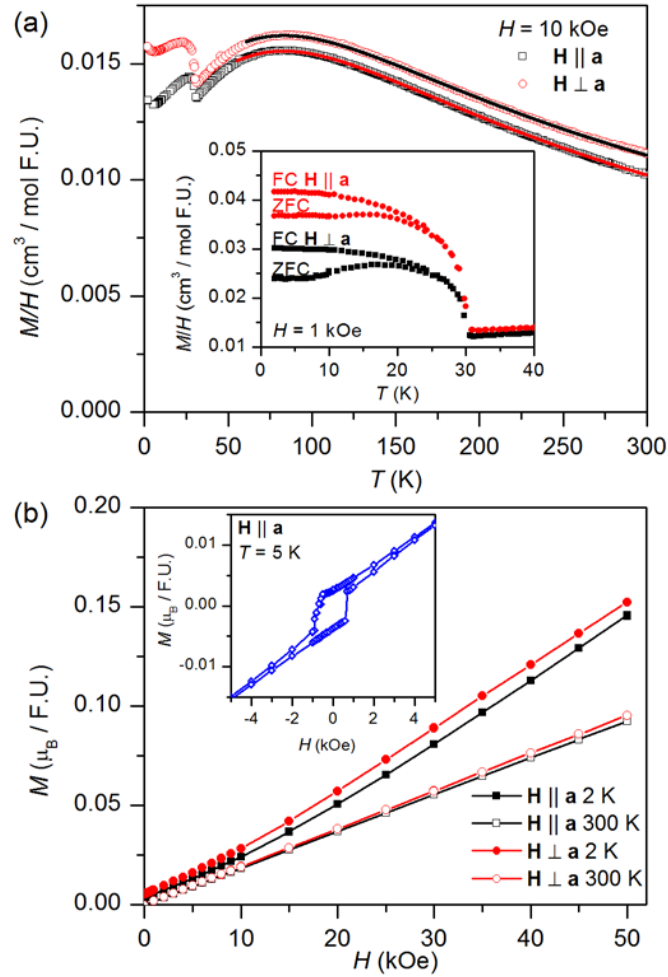

Figure 3. (a) Magnetic susceptibility M/H as a function of temperature at $H = 10$ kOe, featuring a broad hump near 80 K characteristic to the onset of 1D short-range correlations, and a sharp feature indicating long range magnetic ordering below 30 K. The inset in (a) displays the divergence between FC and ZFC data measured at $H = 1$ kOe suggesting a canted spin configuration in $SrMn(VO_4)(OH)$. (b) Isothermal magnetization curves at 2 K and 300 K showing a small remnant magnetization and a field induced spin reorientation at low temperatures. The inset in (b) shows the magnetic hysteresis in the long range ordered state.

The fits of the parallel and perpendicular susceptibility data for temperatures above 55 K are shown as lines in Figure 3a and give very similar results: $J/k_B = -9.97 \pm 0.03$ K and $g = 2.02 \pm 0.02$. The temperature independent term $\chi_0$ is only needed for fitting the perpendicular susceptibility, which yields a value of $0.0010 \pm 0.0006$ cm$^3$/mol FU. This probably arises from failure to completely account for the sample holder contribution to the measured magnetization, but it could indicate a small and temperature independent paramagnetic anisotropy of about 8%. Among the other Mn$^{2+}$-based 1D antiferromagnets, CsMnBr$_3$ appears to have the closest resemblance to SrMn(VO$_4$)OH in the terms of interaction strength along the chain ($J/k = 9.9(4)$) [9].

Heat capacity measurements were also performed to further probe the magnetic transition near 30 K. The results are shown in Figure 4. Measurements in zero magnetic field show a sharp anomaly confirming the bulk nature of the phase transition at $T_N = 30$ K. Application of a 50 kOe field is observed to produce only a small shift of 0.4 K in the heat capacity peak position, but broadens the transition toward higher temperatures. This is consistent with a small ferromagnetic component associated with the canted structure. An estimate of the magnetic entropy associated with the phase transition gives 0.075 R per mole of Mn. This relatively small value is consistent with a transition from a partially (1d) ordered state to full long range order in 3 dimensions.

The low temperature heat capacity is shown in the inset of Figure 4. It is plotted as $c_P/T$ vs $T^2$, and the linear behavior shows that the data is well described by $c_P(T) = \beta T^3$, that is, the low temperature limit of the Debye function representing the phonon or lattice contribution [36]. No electronic contribution is expected in this insulating material. The fit shown in the inset is a line forced to pass through the origin. The fitted $\beta$ is $2.29(1) \times 10^{-5}$ R/mol-at./K$^3$. In these units, the Debye temperature $\Theta_D$ can be determined from the relationship $\Theta_D = (234 / \beta)^{1/3}$, giving a Debye temperature of 217 K. The analyzed data are collected below about 0.2 $T_N$, so magnetic contributions to the heat capacity are neglected. We note that if present, the magnetic contribution in this antiferromagnet would be expected to have a $T^3$ temperature dependence, and be indistinguishable from the lattice heat capacity.[37]

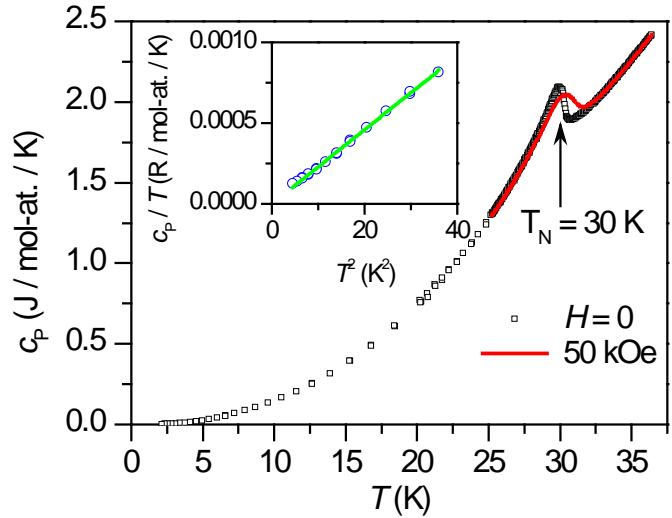

Figure 4. Heat capacity ($c_P$) measured at H = 0 and 50 kOe showing a sharp anomaly at the magnetic ordering temperature. The inset in (b) show the linear behavior of $c_P/T$ vs $T^2$ at low temperature

### *D. Magnetic structure of SrMn(VO$_4$)(OH)*

Neutron diffraction measurements were performed above and below the magnetic ordering temperature, $T_N$ = 30 K. No indication of structural transformation has been observed within the HB3A wave vector resolution. Refinements were carried out using the structural model obtained from x-ray diffraction data, and R-factors ($R_F$) of 5.23 % and 4.87 % were obtained for 40 K and 4 K data, respectively. The refined structural parameters are very similar to those obtained from the x-ray data, listed in Table I. A comparison of the observed and calculated intensities for 40 K and 4K is shown in Figure 5. Note that the intensities measured at 4 K contain both nuclear and magnetic scattering contributions. For probing the nuclear structure below the magnetic transition one have to consider only the Bragg reflections at relatively large momentum transfer (Q > 4 Å$^{-1}$) where the magnetic form factor suppresses the magnetic intensity almost completely. The estimated magnetic contribution to the 4 K data is depicted by black columns in the lower panel of Fig. 5. The large Q Bragg peaks intensities remain almost unchanged within the experimental error.

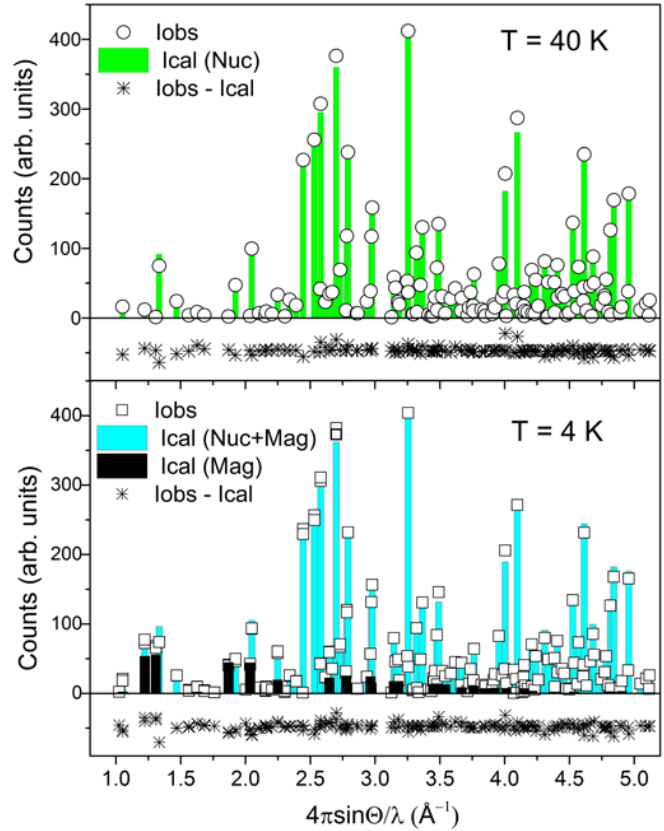

Figure 5. Comparison of the observed and calculated neutron scattering intensities of the SrMn(VO$_4$)(OH) in the disordered state at 40 K, and in the antiferromagnetically ordered state at 4 K. The Bragg intensities measured at 4 K consist of both nuclear and magnetic scattering contributions and the estimated magnetic contribution is depicted by black columns.

A comprehensive survey of the reciprocal lattice at 4 K revealed a magnetic order with the propagation vector **k** = (0, 0, 0). The scattering intensity of the (101) Bragg peak has been measured as a function of temperature and is shown in Figure 6. To determine the character of the long-range-order transition, we evaluated the critical exponent $\beta$ associated with the magnetic order parameter defined as $m(T) \propto \sqrt{I} \propto (T_N - T)^\beta$. Least-square fits were carried out for different temperature intervals around transition ($\Delta T = T_N - T$). The fitted value of $T_N = 30.0(5)$ K is in good agreement of the magnetization measurements, and the obtained $\beta$ values are plotted in the insert of Figure 6 as a function of $\Delta T$. The estimated value of the critical exponent $\beta$ for $\Delta T \rightarrow 0$ is 0.32(1), very close to the value expected for the 3D Ising model ($\beta = 0.3265(3)$) [38]. We note that similar Ising critical exponent has recently been reported for the S=3/2 quasi-one-dimensional antiferromagnet $SrCo_2V_2O_8$, while the order parameter for the S=5/2 congener $SrMn_2V_2O_8$ has been best described by a 3D Heisenberg exponent[22].

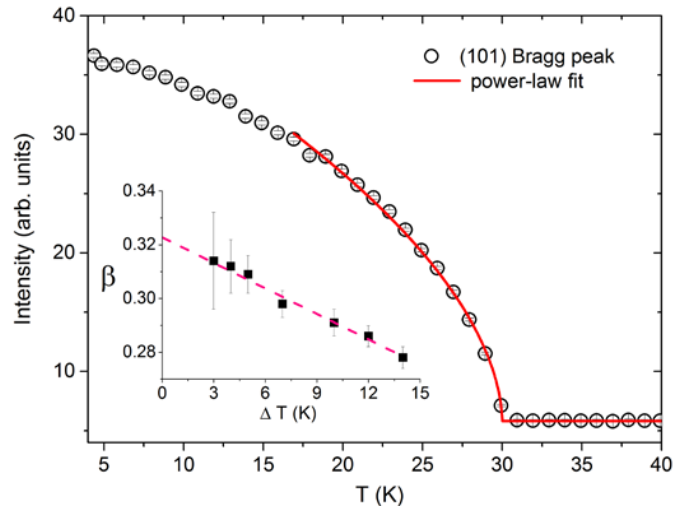

Figure 6. Temperature dependence of the (101) Bragg peak intensity and the fit using a power-law model, as described in the text. The critical exponent $\beta$ associated with the magnetic order parameter is plot in the in the insert.as a function of fitting range ($\Delta T$).

The single crystal magnetic structure at 4 K has been determined from a set of 181 unique reflections. The magnetic contribution has been first estimated from the difference of scattering intensity between 4 K and 40 K. Representational analysis has been employed to determine of the symmetry-allowed magnetic structures that can result from a second-order magnetic phase transition, given the $P2_12_12_1$ crystal structure before the transition and the propagation vector **k** =

(0,0,0). These calculations were carried out using version 2K of the program SARAh Representational Analysis [39]. The decomposition of the magnetic representation $\Gamma_{mag}$ for the Mn site in terms of the non-zero irreducible representations (IRs) and their associated basis vectors, $\psi$, is given in Table 3. The labeling of the propagation vector and the IRs follows the scheme used by Kovalev [40]. By trying several models, it was found that using basis vectors of a single IR fails to give a satisfactory fit to the observed intensities. The best fit is obtained by combining the basis vector $\Psi_2$ and $\Psi_3$ of representations $\Gamma_1$ (magnetic space group $P2_12_12_1$) and $\Gamma_2$ ($P2_1'2_1'2_1$), respectively. These basis vectors enable antiferromagnetic orientation of the nearest neighbor spins both along and between the chains. Note that such a magnetic order does not follow the Landau's theory of second-order phase transitions. In this context it is informative to perform a symmetry characterization in the form of a symmetry group that allows understanding the magneto-structural implications [41]. For this we used the tools available at the Bilbao Crystallographic Server [42,43]. We determined that the actual symmetry of the magnetic structure of SrMn(VO$_4$)(OH) is given by a monoclinic subgroup of type $P2_1$. In this case the Mn site (Wyckoff *4a* in $P2_12_12_1$) is split into two independent sites, each defining a different [MnO$_4$]$_\infty$ chain running along the *a*-direction. The structural degrees of freedom induced by the magnetic ordering are, however, negligible within our experimental resolution. In the parent-like lattice setting, the 2*a* Wyckoff position of the $P2_1$ magnetic space group consists of (x, y, z; m$_a$, m$_b$, m$_c$) and (x+1/2, -y+1/2, z; m$_a$, -m$_b$, -m$_c$). The lowering of symmetry to the polar point group 2 allows one to infer that this system could exhibit multiferroicity, with a magnetically induced ferroelectricity.

**Table 3.** Basis vectors for the space group $P2_12_12_1$ with propagation vector **k** = (0, 0, 0). The decomposition of the magnetic representation for the Mn site is $\Gamma_{mag}= 3\,\Gamma_1 + 3\,\Gamma_2 + 3\,\Gamma_3 + 3\,\Gamma_4$. The atoms four equivalent Mn positions are defined as 1: (0.255, 0.256, 0.995), 2: (0.755, 0.243, 0.004), 3: (0.744, 0.756, 0.504), 4: (0.244, 0.743, 0.495).

| IR | Atom | $\Psi1$ | $\Psi2$ | $\Psi3$ | Magn. SG |
|---|---|---|---|---|---|
| $\Gamma_1$ | 1 | ( 1  0  0) | ( 0  1  0) | ( 0  0  1) | $P2_12_12_1$ |
|  | 2 | ( 1  0  0) | ( 0 -1  0) | ( 0  0 -1) |  |
|  | 3 | (-1  0  0) | ( 0  1  0) | ( 0  0 -1) |  |
|  | 4 | (-1  0  0) | ( 0 -1  0) | ( 0  0  1) |  |
| $\Gamma_2$ | 1 | ( 1  0  0) | ( 0  1  0) | ( 0  0  1) | $P2_12_1'2_1'$ |
|  | 2 | ( 1  0  0) | ( 0 -1  0) | ( 0  0 -1) |  |
|  | 3 | ( 1  0  0) | ( 0 -1  0) | ( 0  0  1) |  |
|  | 4 | ( 1  0  0) | ( 0  1  0) | ( 0  0 -1) |  |
| $\Gamma_3$ | 1 | ( 1  0  0) | ( 0  1  0) | ( 0  0  1) | $P2_1'2_12_1'$ |
|  | 2 | (-1  0  0) | ( 0  1  0) | ( 0  0  1) |  |
|  | 3 | (-1  0  0) | ( 0  1  0) | ( 0  0 -1) |  |
|  | 4 | ( 1  0  0) | ( 0  1  0) | ( 0  0 -1) |  |
| $\Gamma_4$ | 1 | ( 1  0  0) | ( 0  1  0) | ( 0  0  1) | $P2_1'2_1'2_1$ |
|  | 2 | (-1  0  0) | ( 0  1  0) | ( 0  0  1) |  |
|  | 3 | ( 1  0  0) | ( 0 -1  0) | ( 0  0  1) |  |
|  | 4 | (-1  0  0) | ( 0 -1  0) | ( 0  0  1) |  |

Referring to the atom positions labeled in Table 3, the full magnetic configuration can be described as: 1: ($m_a$, $m_b$, $m_c$), 2: ($m_a$, -$m_b$, -$m_c$), 3: ($m_a$, $m_b$, $m_c$), 4: ($m_a$, -$m_b$, -$m_c$). In addition to the alternating sign of the *b* and *c* moment projections, it can be noticed that the $P2_1$ magnetic space group allows for a ferromagnetic component along the *a-* axis. Such a canted antiferromagnetic structure can be achieved through the Dzyaloshinskii-Moriya interaction, which minimizes the exchange energy when the spins are orthogonal. The magnetic structure is depicted in Figure 7.

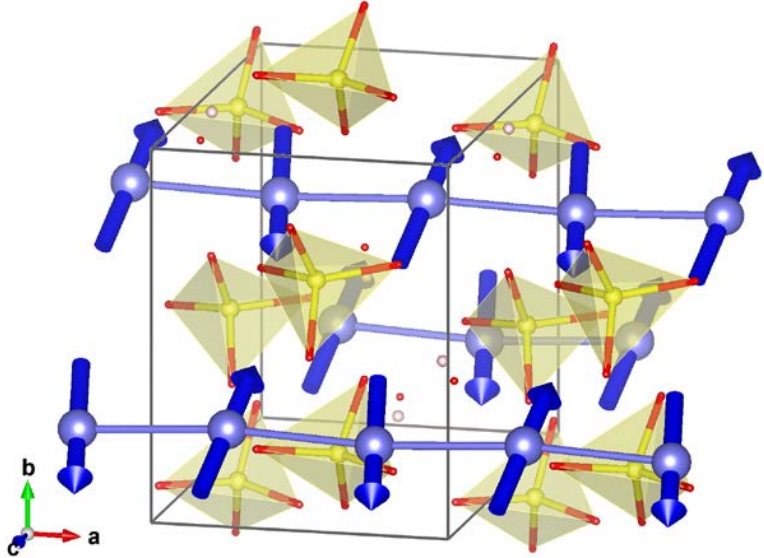

Figure 7. Magnetic structure of SrMn(VO$_4$)(OH) at 4 K. Only the Mn and VO$_4$ tetrahedra are shown for clarity. The nearest neighbor Mn moments are ordered antiferromagnetically in the *bc* plane and are slightly canted along the *a*-direction.

The final least-squares refinement of the magnetic structure was performed using combined magnetic and nuclear phases contributions to the measured reflections at 4 K. The extinction coefficients and thermal parameters were fixed to the values obtained from the paramagnetic state, measured at 40 K. The magnitude of the moment was constrained to be the same for all Mn atoms. The refinement including both magnetic and nuclear contributions to the scattering yielded a R$_F$ factor of 4.87 %. The refined moment projections are m$_a$ = 0.5(2) µ$_B$, m$_b$ = 2.0(2) µ$_B$ and m$_c$ = -2.6(2) µ$_B$. The moments are nearly parallel to the [011] direction and their coupling along the chain as well as between adjacent chains (connected through a *tet*-VO$_4$ unit) is predominately antiferromagnetic. The additional ferromagnetic component along the chain direction is small and subject to a larger experimental error. The canting angle estimated from such a moment projection is 8 +/- 3 degrees. The amplitude of the total static moment is 3.4(1) µ$_B$, significantly smaller than the fully ordered moment of 5 $\mu B$ (Mn$^{2+}$: *S* = 5/2). Since the order parameter shown in Figure 6 indicates that the ordered moment should be close to saturation at 4

K, a reduced moment could indicate the persistence of strong spin fluctuations due to the quasi-1D nature of magnetic interactions.

## IV. CONCLUSIONS

Single crystals of the formula SrMn(VO$_4$)(OH) have been grown using a high temperature hydrothermal method. The structure is reminiscent of the adelite-descloizite mineral family and contains linear S=5/2 Mn$^{2+}$ ions in octahedral environments coupled by tetrahedral vanadate linkages. Single crystal X-ray diffraction provides the details of the crystal structure and confirms that it is in the non-centrosymmetric $P2_12_12_1$ space group. Vibrational spectroscopy analysis is typical of bridging vanadates and also confirms the presence of coordinated OH$^-$ groups. The synthesis method enables the growth of multi-millimeter single crystals enabling the measurement of single crystal magnetic and single crystal neutron diffraction measurements. The magnetic measurements taken over a range of fields and temperatures display a broad maximum at higher temperatures suggesting short-range 1D coupling. Fitting with a modified Bonner-Fisher model describing the intrachain interactions between the S=5/2 spins on Mn$^{2+}$ ions gives $J/k_B = -9.97 \pm 0.03$ K and $g = 2.02 \pm 0.02$. Neutron diffractions shows that below 30 K antiferromagnetic long-range order develops with a small canting, which is likely a result of the Dzyaloshinskii-Moriya interaction. The magnetic order involves a reduction in symmetry described by a monoclinic magnetic space group P2$_1$, and consists of two unique magnetic chains with antiferromagnetic nearest-neighbor exchange couplings. The magnetic phase transition is observed in both magnetic susceptibility and heat capacity measurements. The canting results in a small ferromagnetic and hysteretic component in the low temperature magnetization. The physical properties of this system suggest a variety of complex magnetic and structural interactions over the coupled one-dimensional chains. The structural variety of the tetrahedral bridging group provides the ability to systematically examine a wide range of linear chains with high spin states.

## Acknowledgments:

The authors thank the National Science Foundation for financial support under Grants No. DMR-1410727 and 1307740 for financial support. Work at the Oak Ridge National Laboratory was sponsored by the US Department of Energy, Office of Science, Basic Energy Sciences, Scientific User Facilities Division (neutron diffraction) and Materials Sciences and Engineering Division (physical property measurements).